\documentstyle[twoside,fleqn,espcrc2,psfig]{article}

\newcommand{\AmS}{{\protect\the\textfont2
  A\kern-.1667em\lower.5ex\hbox{M}\kern-.125emS}}
\title{Thermodynamics of 1-flavor QCD
\thanks{Talk presented by Ph. de Forcrand at LATTICE96}}

\author{Constantia Alexandrou\address{Department of Natural Sciences, 
        University of Cyprus, CY-1678 Nicosia, Cyprus},
        Artan Bori\c{c}i\address{Swiss Center for Scientific Computing, 
        ETH-Zentrum, CH-8092 Z\"urich, Switzerland},
        Alessandra Feo$^{\rm a}$,
        Philippe de Forcrand$^{\rm b}$,
        Andrea Galli\address{ELCA Informatique, HofwiesenStr. 26, 
        CH-8057 Z\"urich, Switzerland},
        Fred Jegerlehner\address{DESY-IfH Zeuthen, D-15738 Zeuthen, Germany},
        and 
        Tetsuya Takaishi$^{\rm b}$
}

\begin{document}

\begin{abstract}
We present first results, for heavy to moderate quark masses, of a study
of thermodynamic properties of 1-flavor QCD, using the multiboson algorithm.
\end{abstract}

\maketitle

\section{Introduction}
Our understanding of QCD thermodynamics is still very incomplete. 
Simulations of the 0-flavor (quenched) lattice theory are just starting
to produce results in line with continuum expectations. The situation
in the 2- and 3-flavor cases is still rather murky (for an excellent review,
see \cite{Kanaya}, or this volume). Additional information for different
numbers $n_f$ of quark flavors is clearly desirable to build a coherent 
picture such as proposed in \cite{PW} near the chiral limit. The study
of QCD with exotic $n_f$ has focused on many flavors \cite{Iwasaki}. 
Perhaps because of algorithmic difficulties, 1-flavor QCD has been 
largely ignored, although it has particularly interesting properties.

It is impossible to make a $\pi - \sigma$ doublet with only 1 quark species,
so there is no chiral symmetry to speak of at vanishing quark mass.
The low-lying meson spectrum consists of the ``$f_0$'' ($\overline{q} q$) and
the ``$\eta'$'' ($\overline{q} \gamma_5 q$). The $\eta'$ correlator
$G_{\eta'}(x,y) = |D^{-1}(x,y)|^2 - D^{-1}(x,x) D^{-1}(y,y)$ contains
2 pieces, only the first of which, analogous to the pion correlator,
becomes massless (constant) at $m_q = 0$. Thus the chiral phase transition
and the associated Goldstone boson are absent in the 1-flavor theory
\cite{PW}. In the heavy quark regime, on the other hand, the effect of
dynamical quarks is still analogous to that of an external magnetic
field on the equivalent Potts system. Just like in the 2-flavor case,
this external field weakens the phase transition, 
possibly turning it into a crossover. Thus, as the quark mass is decreased,
one might expect the first-order deconfinement transition to become
second-order, with an end-point at some finite quark mass. For any 
lighter quark, only a crossover would be seen.

Of course, testing this scenario requires careful finite-size scaling
studies. As a first step, we present here preliminary results obtained 
on an $8^3 \times 4$ lattice. We use Wilson fermions, and try to follow
the critical line in the $(\kappa, \beta)$ plane, starting from $\kappa = 0$.

\section{Algorithm}
The standard algorithm to study any number of flavors is the R-algorithm
\cite{Sugar}, where the effect of the fermionic determinant ${\rm det} D$ is 
replaced 
by discretized Brownian noise. This discretization entails step-size errors,
so that a delicate and costly extrapolation to zero step-size must be performed.
Instead, we use here the multiboson method \cite{Luscher,Galli}. In its
non-hermitian variant, it can simulate 1-flavor QCD \cite{Borici}.

Approximating $D^{-1}$ by a Chebyshev polynomial 
$P_{2n}(D) = c_{2n} \prod_{k=1}^n (D - z_k) (D - \overline{z_k})$,
and noticing that 
${\rm det}(D - \overline{z_k}) = {\rm det} \gamma_5 (D -
\overline{z_k}) 
\gamma_5 = {\rm det}(D - z_k)^\dagger$,
one obtains
\begin{equation}\label{Tn}
{\rm det} D \approx {\rm det}^{-1} P_{2n}(D) = | {\rm det} T_n(D) |^{-2}
\end{equation}
with
\begin{equation}
T_n(D) = \sqrt{c_{2n}} \prod_{k=1}^n (D - z_k)
\end{equation}
The right-hand side of (\ref{Tn}) can then be replaced by a Gaussian integral
over $n$ bosonic fields $\phi_k$. The action is
$S_{\rm tot} = S_{\rm gauge} + \sum_{k=1}^n |(D - z_k) \phi_k|^2$.
Because (\ref{Tn}) is an approximation, a corrective Metropolis test is added,
just like in the 2-flavor algorithm \cite{Galli}. At the end of each
``trajectory'' $\{U,\phi\} \rightarrow \{U',\phi'\}$, the configuration 
is accepted with probability $min(1,e^{- \eta^\dagger (W^\dagger W - 1) \eta})$,
where $W = [T_m(D') T_n^{-1}(D')] [T_m(D) T_n^{-1}(D)]^{-1}$.
This algorithm samples the distribution 
$e^{-S_{\rm gauge}} {\rm det}^{-1} P_{2m}(D)$.
With negligible overhead, $m$ can be taken very large ($m \geq 3 n$ in our case),
so that the measure is effectively $e^{-S_{\rm gauge}} {\rm det}^{-1} 
P_\infty(D)$.

Now the domain of convergence ${\cal S}$ of the Chebyshev approximation
$P_{2m}(z) \approx 1/z$ in the complex plane is bounded by an ellipse 
$\partial {\cal S}$ centered at $(1,0)$ which goes through the origin.
If $z \in {\cal S}$, $z P_{2m}(z) = 1 + {\cal O}(e^{-\alpha m})$:
the approximation is exponentially convergent. Otherwise, 
$|z P_{2m}(z)| \rightarrow \infty$ as $m \rightarrow \infty$. 
This means that configurations for which eigenvalues of $D$ fall outside
${\cal S}$ will always be rejected by our algorithm. The measure we sample
is $e^{-S_{\rm gauge}} {\rm det}D~~\Theta({\rm spectrum}(D) \in {\cal S})$,
where $\Theta(u) = 1$ or 0 if $u$ is true or false.

We take for $\partial {\cal S}$ a circle, which contains the spectrum of
$D = {\bf 1} - \kappa M$ 
at $\beta = \infty$ for any $\kappa < \kappa_c = 1/8$,
and at $\beta = 0$ for any $\kappa < \kappa_c = 1/4$ in the thermodynamic 
limit. Thus we expect it to contain the spectrum of $D$ also at intermediate
$\beta$ for $\kappa < \kappa_c(\beta)$, in the thermodynamic limit.
Finite-volume fluctuations cause some ``exceptional configurations'',
with eigenvalues outside ${\cal S}$, to be included in the exact partition
function. Among those are configurations with negative real eigenvalues,
for which ${\rm det} D < 0$: they create the notorious ``sign problem''
associated with non-degenerate Wilson quarks. Our algorithm discards
such configurations 
\footnote{Note that the R-algorithm would include them in the partition
function, but with the wrong sign.}.

Therefore, even though it recovers the correct measure in the thermodynamic
limit, our algorithm introduces a bias on a finite lattice. This bias 
however can be viewed as a finite-size effect: it only appears if the
inverse quark mass becomes too large compared with the lattice size.
For the relatively heavy quarks we consider here, this bias is completely
negligible.

The implementation of our algorithm on the Cray-T3D is very similar to the
2-flavor case \cite{Galli}. Here too, even-odd preconditioning can be
implemented through a simple redefinition of the zeroes $z_k$. Half as
many fields $\phi_k$ are needed for the same quality of the approximation
(\ref{Tn}). The Metropolis acceptance is shown in Fig.1, to be compared
with the 2-flavor case.
\begin{figure}[htb]
\vspace{-0.7cm}\psfig{figure=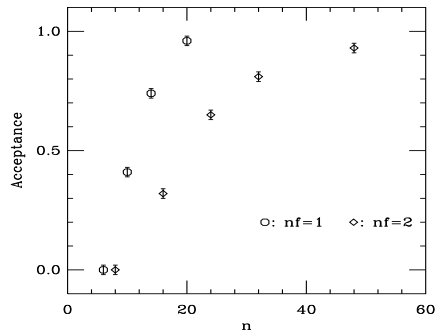,height=6cm,width=7.5cm}\vspace{-2.3cm}
\caption{Acceptance versus number of auxiliary fields for $4^4, 
\beta=0,k=0.215$. Circles are $n_f=1$ data. Diamonds are $n_f=2$ data.}
\end{figure}
\vspace{-1cm}

\section{Results}

For a fixed $\kappa$, we tried to scan $\beta$ and identify ``$\beta_c(\kappa)$''.
On our rather small lattice, different criteria for criticality will give
slightly different effective $\beta_c$'s. We tried to choose a criterion
which reduced finite-size effects, at least in the quenched case.
Susceptibility measurements, of the Polyakov loop especially, had large
statistical errors and were difficult to exploit.
The most useful observables for us were:\\
i) the histogram of the norm of the Polyakov loop, which shows 2 peaks near
criticality. This is safe only if many tunneling events occur between the
2 phases. Fortunately that is the case with our algorithm, as shown in Fig.2
for $\kappa = 0.10$. The work per sweep is roughly equivalent to 6 molecular
dynamics steps of the R-algorithm, which shows that the multiboson approach
is rather efficient to decorrelate the Polyakov loop, a global observable.\\
\begin{figure}[h]\vspace{-2.5cm}
\psfig{figure=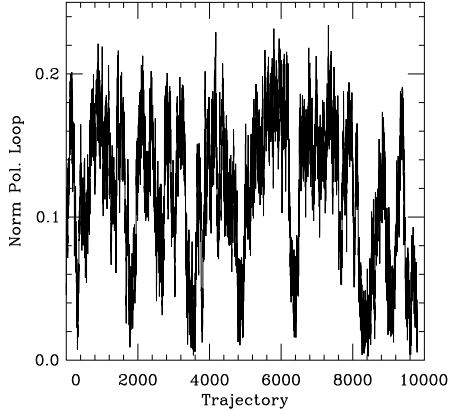,height=7.5cm,width=7.5cm,clip=}\vspace{-1.3cm}
\caption{History of the norm of the Polyakov loop for $\beta=5.67,k=0.1, L=8^3\times 4$. 
The work per trajectory is equivalent to $\sim 6$ molecular dynamics {\em steps}.}
\end{figure}
\vspace{-0.8cm}
ii) the deconfinement ratio $\rho \equiv 3/2 p - 1/2$, where $p$ is the
probability for the complex trace of the Polyakov loop to fall within 
$\pm 20^{\circ}$ of a $Z_3$ axis. $\rho$ goes from 0 to 1 as the $Z_3$ 
symmetry is broken. A preliminary quenched study showed us that identifying
$\beta_c$ as that for which $\rho = 75 - 80\%$ gave very small finite-size
corrections.

We then used these 2 observables to determine $\beta_c$ for 
$\kappa = 0.05, 0.10, 0.12, 0.14$. The number of auxiliary fields $\phi_k$ 
was $8, 16, 24, 32$ respectively. Representative histograms of the norm of the Polyakov
loop are shown in Fig.3, to compare with the quenched case. A 2-peak structure remains visible up to the
highest $\kappa$. One can also clearly see the breaking of the $Z_3$ 
symmetry by the dynamical quarks: the peak corresponding to the confined
(disordered) phase moves to the right as $\kappa$ increases. This is
evidence for the expected weakening of the phase transition, which should
be confirmed by simulations on larger spatial volumes and by finite-size
scaling.

Based on Fig.3 and on the measurement of the deconfinement ratio,
our estimates for 
$\beta_c(\kappa)$ are given in Table 1, together with published results
for $n_f = 0$ and 2. One sees little departure so far from a linear behavior
in $n_f$: the ordering effect of 1 flavor of quarks is about $1/2$ that
of 2 flavors.

{\small
\begin{center}
\begin{tabular}{|c|c|c|}
\hline
& \multicolumn{2}{c|}{$N_f$} \\ \cline{2-3}
\multicolumn{1}{|c|}{$\kappa$}
& 1 & 2 \\ \hline
0    & \multicolumn{2}{c|}{5.69254(24)} \\ \hline
0.05 & 5.685(5)  & - \\ \hline
0.10 & 5.67(1)   & - \\ \hline
0.12 & 5.630(15) & 5.58(2) \\ \hline
0.14 & 5.59(2)   & 5.46(2) \\ \hline
\end{tabular}
\end{center}
Table 1: {$\protect\beta_c(\kappa)$ for the quenched \protect\cite{QCDPAX}, 
one- and two-\protect\cite{HEMCGC} flavor theories}

\nopagebreak
\vspace{-2cm}
\begin{figure}[bh]\vspace{-0.5cm}
\centerline{\psfig{file=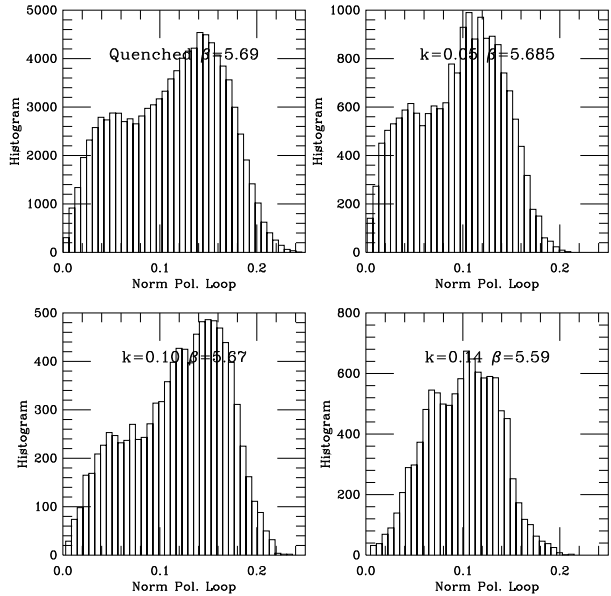,width=11cm,height=11cm,clip=}}\vspace{-1.9cm}
\caption{Histograms of the norm of the Polyakov loop for parameters near $\beta_c$.}
\end{figure}

\end{document}